\documentclass[showpacs,preprintnumbers, amsmath,amssymb,superscriptaddress,preprint]{revtex4}
\usepackage{graphicx}
\begin{document}
\newcommand {\bb}{\bibitem}
\newcommand {\be}{\begin{equation}}
\newcommand {\ee}{\end{equation}}
\newcommand {\bea}{\begin{eqnarray}}
\newcommand {\eea}{\end{eqnarray}}
\newcommand {\nn}{\nonumber}

\title{High-T$_{c}$ Cuprate Superconductivity in a Nutshell
}

\author{Hyekyung Won}

\address{Department of Physics, Hallym University,
Chuncheon 200-702, South Korea
}

\author{Stephan Haas}
\author{David Parker\footnote{Corresponding author: e-mail: davidspa@usc.edu,
Phone: 213-740-1104}}
\author{Kazumi Maki}

\address{Department of Physics and Astronomy, University of Southern 
California, Los Angeles, CA 90089-0484 USA}

\date{\today}

\begin{abstract}

Since the discovery of high-T$_{c}$ cuprate superconductivity in 1986
many new experimental techniques and theoretical concepts have been developed.
In particular it was shown that the BCS theory of d-wave superconductivity
describes semi-quantitatively the high-T$_{c}$ superconductivity.  Furthermore,
it was demonstrated that Volovik's approach is extremely useful for finding 
the quasiparticle properties in the vortex state. Here we survey these developments 
and forecast future directions.
\end{abstract}
\maketitle

\noindent{\it \bf 1. Introduction}

In 1986 the epoch-making discovery of superconductivity in ceramic La$_{2-x}$Ba$_{x}$CuO$_{4}$
with a transition temperature T$_{c}$ = 35 K by Bednorz and M\"{u}ller \cite{1}
took the scientific community by surprise.  The subsequent enthusiasm as well as the confusion
in the theoretical community are well documented by Charles Enz \cite{2}.

In 1987 P.W. Anderson \cite{3} published his ``dogmas''.  He pointed out that (a) all the action
takes place in the Cu-O$_{2}$ plane common to all high-T$_{c}$ cuprates; and (b) the high-T$_{c}$
superconductivity has to be understood as arising in the middle of a Mott insulator at zero doping.
He proposed a two-dimensional (2D) 1-band Hubbard model as the simplest 
model to describe the high-T$_{c}$ superconductivity.  As a possible ground-state 
wavefunction, he proposed
\bea
\Phi&=& \prod_{i}(1-d_{i})| BCS \rangle\, ,
\eea
where $|BCS\rangle$ is the BCS wave function for s-wave superconductors \cite{4}, 
$d_{i}\equiv n_{i\uparrow} n_{i\downarrow}$ and $\Pi_{i}(1-d_{i})$ 
is the Gutzwiller projector, which annihilates all doubly occupied sites.  
We shall come back to Bob Laughlin's observation on Eq.(1)\cite{5}.

Based on perturbative and numerical analysis of the 2D 1-band Hubbard model a possible d-wave
superconductivity in high-T$_{c}$ cuprates was predicted by a few groups \cite{6,7,8}.  Meanwhile
high-quality single crystals of YBCO, LSCO and thin-film Bi-2212 became available around 1992.
D-wave superconductivity in high-T$_{c}$ cuprates was established finally around 1993-4 through
ARPES \cite{9} and Josephson interferometry \cite{10,11}, among many other experiments.

\noindent{\it \bf 2. BCS Theory of d-wave superconductors with impurities}

Exploring d-wave superconductivity within the BCS framework \cite{12,13,14,15},
we have shown that this theory describes quantitatively the observed quasiparticle density of states
\cite{16} and superfluid density \cite{17} when Zn is substituted for Cu in the Cu-O$_{2}$
plane.

In 1993 Patrick Lee \cite{18} discovered a remarkable phenomenon, i.e. the universal heat conduction.
The quasi-particle spectrum in a d-wave superconductor is given by
\bea
E_{k}&=&\sqrt{v^{2}(k_{\parallel}-k_{F})^{2}+\Delta^{2}\cos^{2}(2\phi)} \\
     &\simeq & \sqrt{v^{2}(k_{\parallel}-k_{F})^{2}+v_{2}^{2}k_{\perp}^{2}}
\eea
with $v_{2}/v = \Delta/E_{F}$ and $k_{\parallel}$ and $k_{\perp}$ the components of ${\bf k}$
parallel and perpendicular to the nodal directions, respectively.  Here the second equation 
is valid in the vicinity of the Dirac cones.

Then the thermal conductivity in the limit $T \rightarrow 0$ and $\Gamma \rightarrow 0$, $\kappa_{00}$
is expressed as
\bea
\kappa_{00}/T&=& \frac{k_{B}^{2}v}{3\hbar v_{2}} n
\eea
where n is the hole or electron density and $E_{F}$ is
the Fermi energy.  Alternatively Eq. 4 can be rewritten as
\bea
\kappa_{00}/\kappa_{n}&=& \frac{2\Gamma}{\pi\Delta}\, ,
\eea
with $\Gamma$ the quasiparticle scattering rate in the normal state, and $\kappa_{n}$ 
the thermal conductivity in the normal state.  May Chiao et al \cite{19,20} then 
deduced $\Delta/E_{F} = \frac{1}{10}$ and $\frac{1}{14}$ for
optimally doped Bi-2212 and YBCO respectively.  Later the thermal conductivity measurement
was extended to LSCO and T1-2210 \cite{21}.  These ratios $\Delta/E_{F}$ imply many things:

a) The pairing in high-T$_{c}$ cuprates is well described by the d-wave BCS theory.  It is far away
from the Bose-Einstein condensation limit. 

b) According to the Ginzburg criterion, the 
fluctuation effects are of order $\sim \Delta/E_{F}$, i.e. they 
can be at most 10 percent.  This appears
to exclude the large phase fluctuation and stripe phase discussed in Refs. \cite{22,23}.

c) For $\Delta/E_{F}= 1/10$ there are hundreds of quasiparticle bound states around
the core of a single vortex in d-wave superconductivity \cite{24,25}.  Indeed the radial ($r$)
dependence of the quasiparticle density of states is very similar to the one obtained for
s-wave superconductivity \cite{26}.  Here $r$ is the distance from the center of the
vortex.  In earlier works \cite{27,28,29,30} it was claimed 
that there would be no bound states around a
single vortex in d-wave superconductivity.  However, in these works it was assumed
that $\Delta \simeq E_{F}$.  It is clear that this unrealistic assumption knocks off most of the
bound states in this analysis.  Unfortunately, these faulty works misled Hussey in
his otherwise excellent review \cite{31} on experiments in high-T$_{c}$ cuprate
superconductors.

In summary, quasiparticles in d-wave superconductors behave as Landau-BCS quasiparticles.
Also the quasiparticles of the normal state are part of a Landau Fermi liquid, although their
properties are rather unorthodox. 

\noindent{\it \bf 3. Semiclassical Approximation}

As discovered by Volovik \cite{32}, the quasiparticle spectrum of the vortex state
in nodal superconductors is calculable within the semiclassical approximation.
Later, Volovik's work was extended for a planar magnetic field \cite{33} and
for thermal conductivity \cite{34}.  Unfortunately, however, Vehkter et al \cite{33}
have used an artificial and unrealistic Fermi surface, while K\"{u}bert et al \cite{34}
have introduced an unphysical spatial average.  These problems were clarified and
corrected in Refs. \cite{35,36,37}.

As is well known the quasiparticle energy in the presence of superflow is given \cite{38} by
\bea
E_{k} \rightarrow E_{k}-{\bf v}\cdot{\bf q}
\eea
Here {\bf v} and 2{\bf q} are the quasiparticle velocity and the pair momentum due
to the superflow.  Also ${\bf v}\cdot{\bf q}$ is known as the Doppler shift (DS).
Then the quasiparticle density of states on the Fermi surface $N({\bf H})$ is given by
\bea
N({\bf H})/N(0) = G({\bf H}) = \langle |{\bf v}\cdot{\bf q}|\rangle/\Delta
\eea
where $\langle \ldots \rangle$ means the average over both the Fermi surface and the vortex lattice.
Then in a magnetic field ${\bf H} \parallel {\bf c}$ in d-wave superconductors,
we obtain
\bea
G({\bf H})&=& \frac{2}{\pi^{2}}\frac{v\sqrt{eH}}{\Delta}
\eea
Here v (= $ 2.6 \times 10^{7}$ cm/sec) is the Fermi velocity within the conduction
plane.  Recent ARPES indicate v is universal and independent of systems (YBCO, LSCO, Bi-2212)
and of doping \cite{9}.  

In all earlier analysis an extra $\pi^{-1}$ factor is missing.
This comes from
\bea
\frac{1}{2\pi}\int_{0}^{2\pi} d\phi \, \delta(\cos(2\phi)) &=& \frac{1}{\pi}
\eea
Actually this factor resolves the long-standing discrepancy between theory and experiment
\cite{39,40}.  When the magnetic field is applied within the a-b plane, we find \cite{35,37}
\bea
G({\bf H}) & \simeq & \frac{2}{\pi^{2}}\frac{\tilde{v}\sqrt{eH}}{\Delta} (0.955+0.0285
\cos(4\phi))
\eea
where $\tilde{v}=\sqrt{v_{c}v}$ with $v_{c}$ the Fermi velocity parallel to the c axis
and $\phi$ is the angle the magnetic field makes from the a axis.
Since the low-temperature specific heat and the spin susceptibility are given by \cite{41}
\bea
C_{s}/\gamma_{N} T &=& G({\bf H}),\,\,\, \chi/\chi_{N}=G({\bf H})
\eea
$G({\bf H})$ should be readily accessible.  However, the $\phi$-angle dependence of $C_{S}$ in
high-T$_{c}$ cuprate superconductors has not been seen yet.

On the other hand, a few thermal conductivity data of optimally doped YBCO have been
reported which exhibit clear fourfold symmetry \cite{42,43,44}.  For $T< (\Gamma\Delta)^{1/2}
<< \tilde{v}\sqrt{eH}$ we obtain \cite{35}
\bea
\frac{\kappa_{xx}}{\kappa_{n}}&=& \frac{\kappa_{yy}}{\kappa_{n}}\simeq \frac{2\tilde{v}^{2}
(eH)}{\pi^{4}\Delta^{2}} (0.955+0.0285\cos(4\phi))^{2}
\eea
Moreover, for $T \gg \tilde{v}\sqrt{eH}$, the sign of the fourfold term becomes negative\cite{37}.
This is observed experimentally in \cite{42,43,44}.

More recently the quasiparticle density of states in the vortex state of a variety
of other nodal superconductors has been analyzed\cite{45,46}.  Using thermal
conductivity measurements, Izawa et al have succeeded
in identifying $\Delta({\bf k})$ of Sr$_{2}$RuO$_{4}$ \cite{47}, CeCoIn$_{5}$\cite{48},
$\kappa$-(ET)$_{2}$Cu(NCS)$_{2}$ \cite{49}, YNi$_{2}$B$_{2}$C \cite{50} and PrOs$_{4}$Sb$_{12}$
\cite{51}.  These  $\Delta({\bf k})$'s are shown in Fig. 1.
\begin{figure}[h]
\includegraphics[width=3.5cm,height=4.5cm]{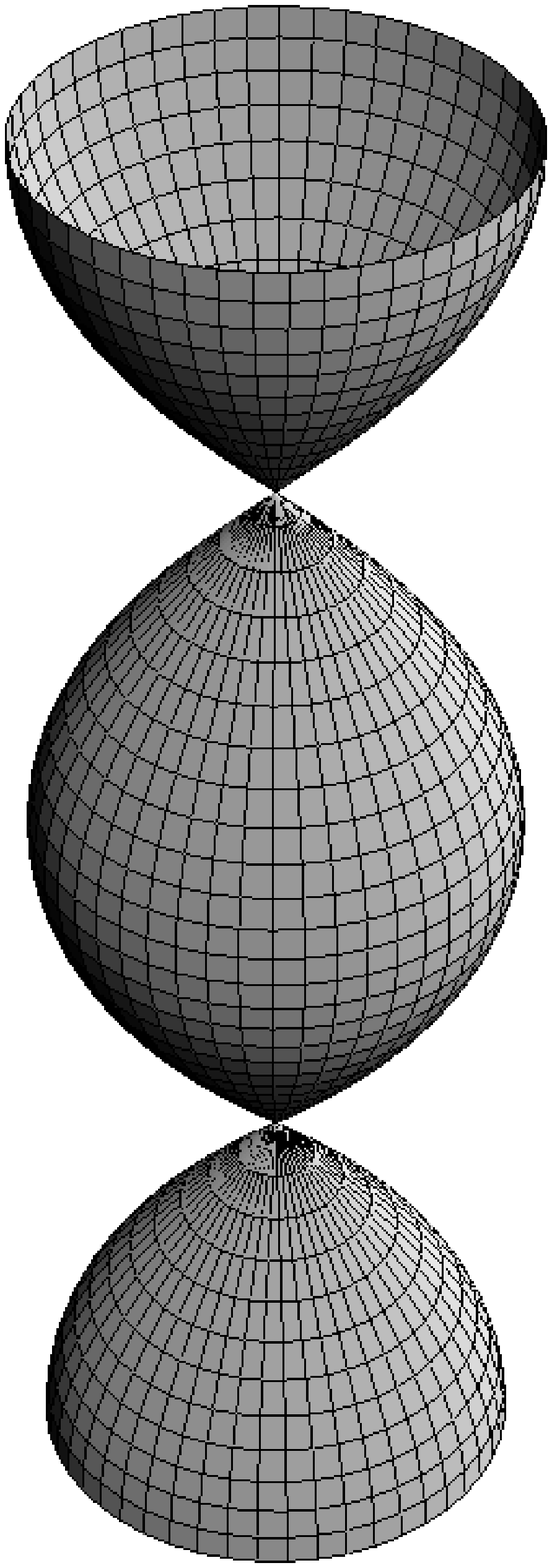}
\includegraphics[width=5.5cm]{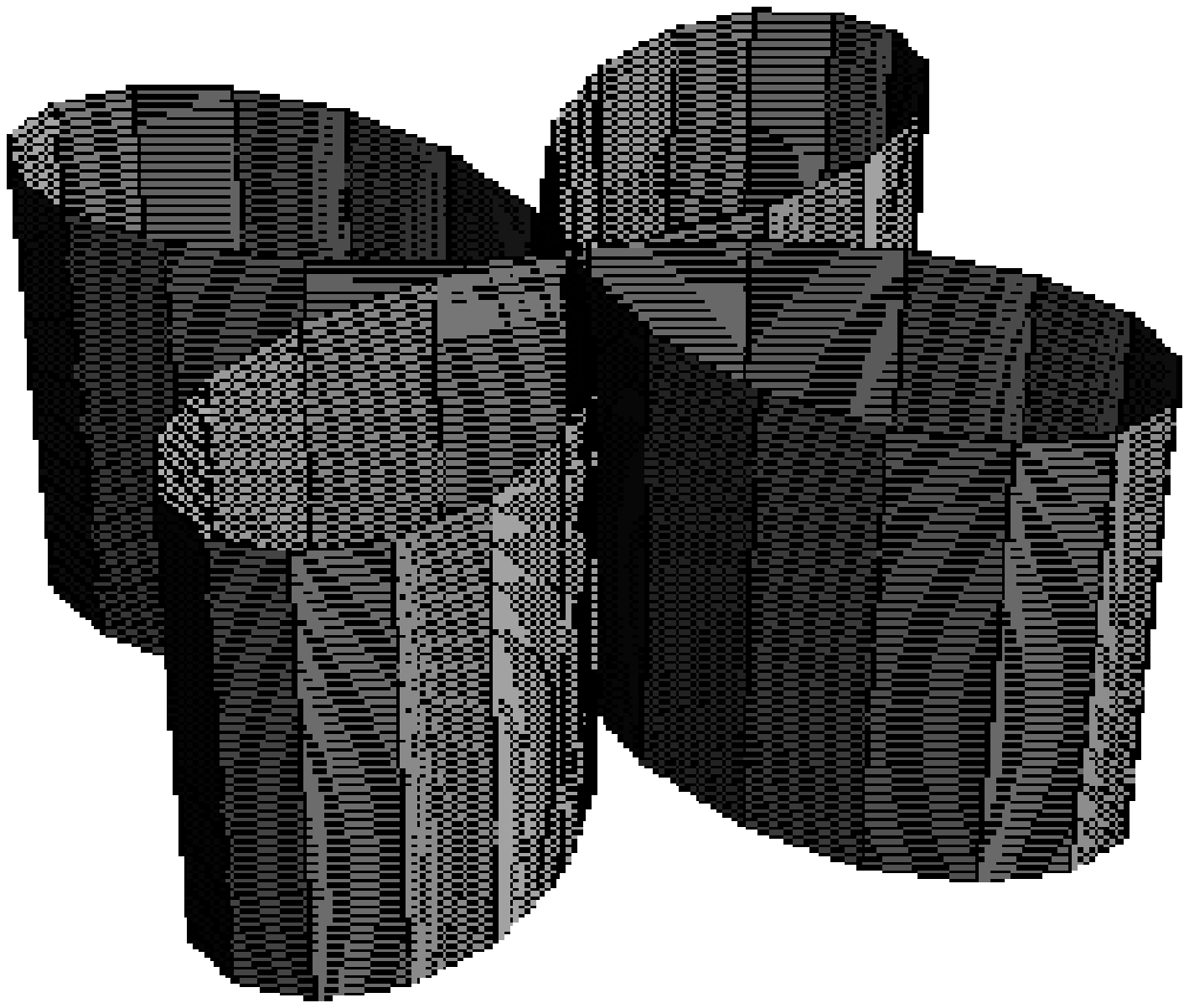}
\includegraphics[width=6.25cm]{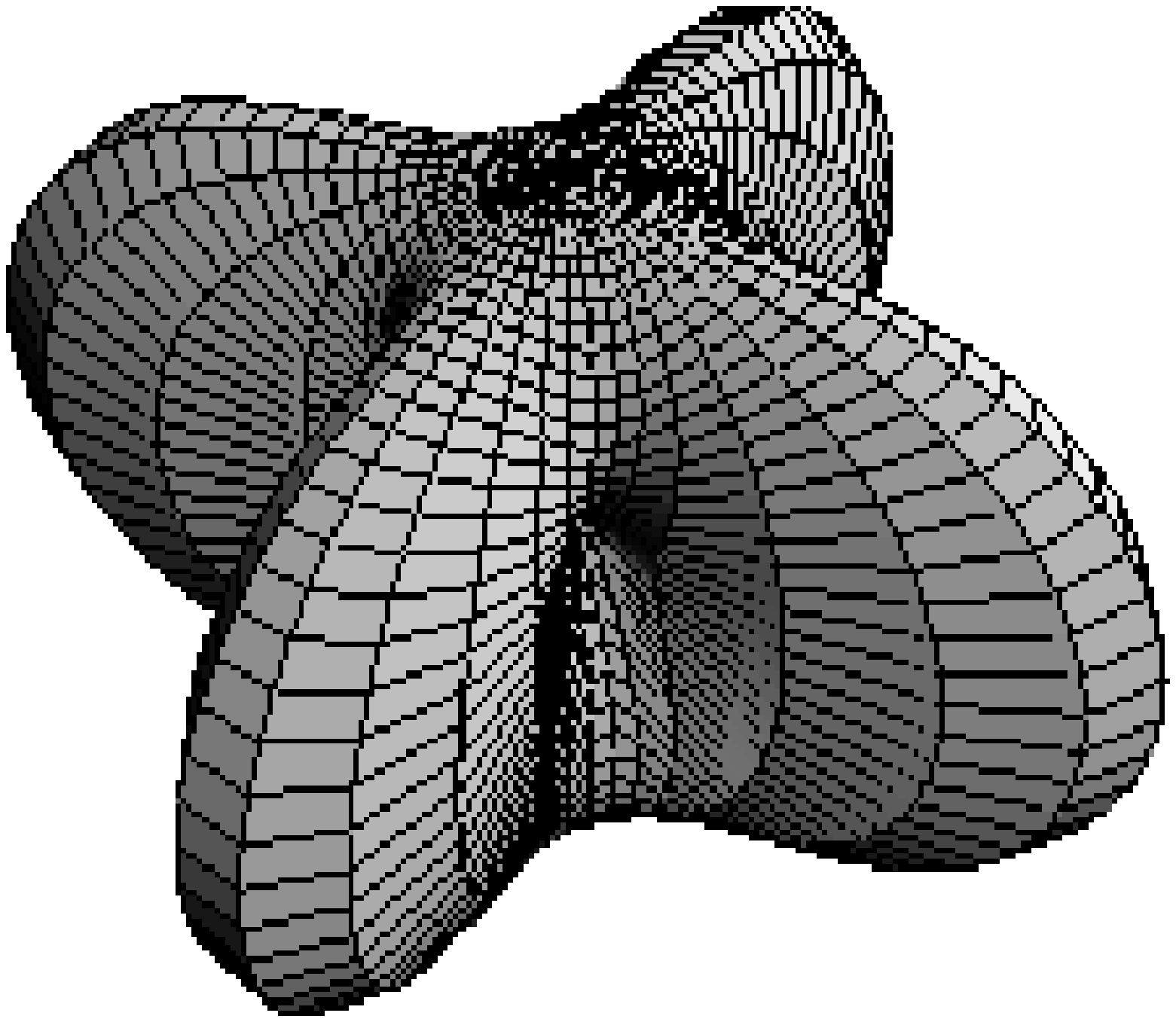}
\includegraphics[width=5cm]{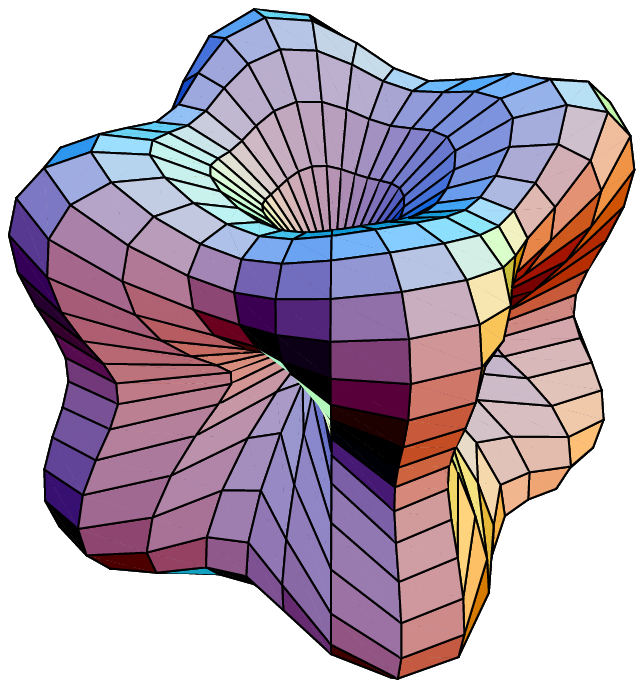}
\includegraphics[width=5cm]{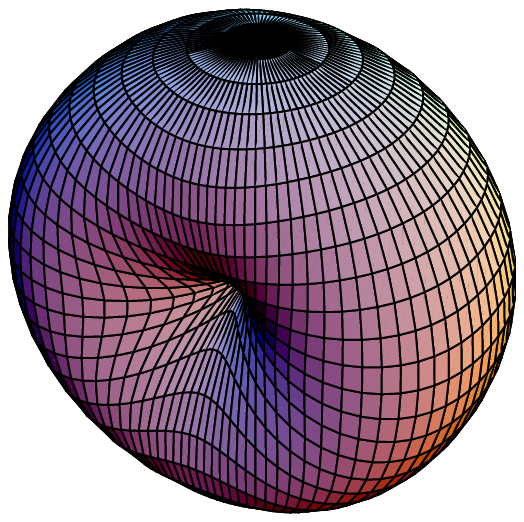}
\caption{From top left, 2D f-wave - Sr$_2$RuO$_4$, d$_{x^{2}-y^{2}}$-wave - CeCoIn$_{5}$ and 
$\kappa$-(ET)$_{2}$Cu(NCS)$_{2}$,s+g-wave - YNi$_{2}$B$_{2}$C, p+h-wave - PrOs$_4$Sb$_{12}$ - 
A phase, p+h-wave - PrOs$_4$Sb$_{12}$ - B phase.}
\end{figure}
In analogy to Eq.(12), the magnetothermal conductivity data in high quality 
single crystals at low temperature
($T \ll \Delta$) provides unique access to the nodal structure of $\Delta({\bf k})$.

\noindent{\it \bf 4. D-wave Density Waves and Gossamer Superconductivity}

In Fig. 2 we sketch the phase diagram of the hole doped high-T$_{c}$ cuprate superconductors.
These materials occupy the region $0.05 < x < 0.25$, where x is the hole concentration.
In the insulating side ($0 \leq x < 0.03$) the antiferromagnetic phase is realized.  The
phase space below $T^{*}$ is called the pseudogap region whose nature has been hotly 
debated.  One may suspect that under the cover of the pseudogap phase
many different phases are hidden.
\begin{figure}[h]
\includegraphics[width=6cm]{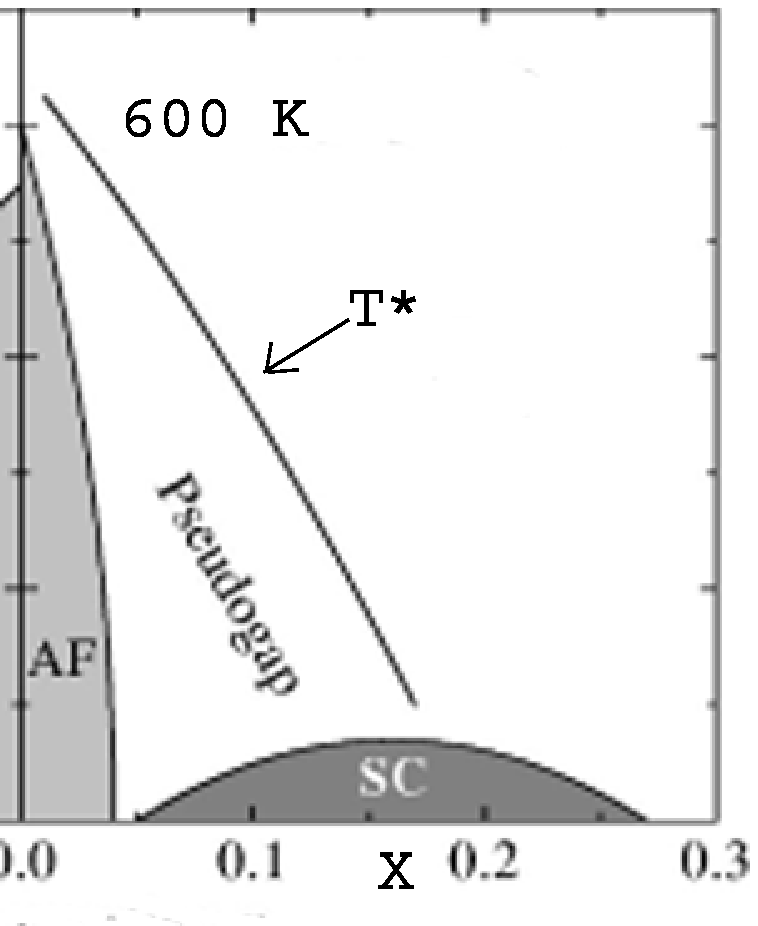}
\caption{Phase diagram of the hole-doped cuprates from Ref. \cite{9}}
\end{figure}
It was proposed \cite{52,53,54} that the pseudogap phase is an unconventional
density wave (UDW).  UDW is a density wave in which the quasiparticle energy gap
has nodes.  Therefore, the transition from the normal state to UDW is a metal-metal
transition, although the quasiparticle density decreases rapidly in UDW.  Furthermore, the local
charge density or the spin density in UDW is hard to observe since $<\Delta({\bf k})>$ =0,
where $<...>$ denotes the average over the Fermi surface.  Therefore UDW is often called a
condensate with a ``hidden order parameter''\cite{53}. For more about UDW we suggest the reader 
to study Ref. \cite{55}.  ARPES data in the pseudogap region 
indicates clearly the d-wave nature of
$\Delta({\bf k})$ \cite{56} (i.e. $\Delta({\bf k}) \sim \cos(2\phi)$).

Although the evidence for d-wave DW or dDW is still elusive, the giant negative
Nernst effect observed in the underdoped region of LSCO, YBCO and Bi-2212 by Wang
et al\cite{57,58} indicates UDW.  It was previously 
shown that the giant negative Nernst effect is the
hallmark of UDW \cite{59}.  Also, UDW appears to describe the observed large Nernst
effect in underdoped LSCO, YBCO and Bi-2212 very consistently \cite{60}.  In the phase
diagram of high-T$_{c}$ cuprates (Fig. 2), it is very likely that 2 order parameters
dDW and dSC (d-wave superconductivity) coexist in the limited region.  This problem has
been briefly discussed in Ref. \cite{53}.  In another paper, Laughlin has presented an intuitive
interpretation of Eq.(1).  The mathematical difficulty of Eq.(1) comes from the
Gutzwiller operator, which has no inverse.  If one replaces $\prod_{i}(1-d_{i})$ by 
$\prod_{i}e^{-\alpha d_{i}}$, Eq.(1) can be understood as a state with competing order
parameters.  The fragile superconductivity in the Mott insulator is called ``gossamer
superconductivity'' \cite{5}.  However, in d-wave superconductivity the Coulomb potential
is not so devastating.  Furthermore, in the region where the superconductivity arises the
antiferromagnetic state has already disappeared.  Instead, the dominant condensate 
is dDW.  Therefore the competition between dDW and 
superconductivity as discussed in \cite{53,61} is more realistic.
In the following we shall use the term ``gossamer superconductivity'' for d-wave superconductivity
in the presence of dDW.

In a pure dDW phase the thermal conductivity exhibits universal heat conduction as long
as imperfect nesting or the chemical potential is neglected \cite{62a}.  This observation is unaffected
in the presence of two competing order parameters 
as long as the imperfect nesting or the 
chemical potential is negligible compared to T or $(\Gamma\Delta)^{1/2}$.  
Then we recover Eq.(3)
\bea
\kappa_{00}/T &=& \frac{k_{B}^{2} E_{F} n}{3\hbar \Delta}
\eea
where $\Delta = \sqrt{\Delta_{1}^{2}+\Delta_{2}^{2}}$ and $\Delta_{1}$ and $\Delta_{2}$ are
the order parameters of dDW and d-wave superconductivity respectively.  In the pseudogap 
region the thermal conductivity can measure $\Delta \simeq \Delta_{1}$.  
This is shown in Fig. 3 (Fig. 6 in \cite{21}).  Clearly $\Delta_{0} \simeq \Delta_{dDW}(0)$
is close to 2.14 $T^{*}$.  Indeed a very similar x dependence of $\Delta$ observed by STM has
been reported \cite{62,63}.  Note that 2.14 is the weak-coupling theory value for both the 
d-wave superconductor \cite{12} and d-wave density-wave \cite{64}.
\begin{figure}[h]
\includegraphics[width=6cm]{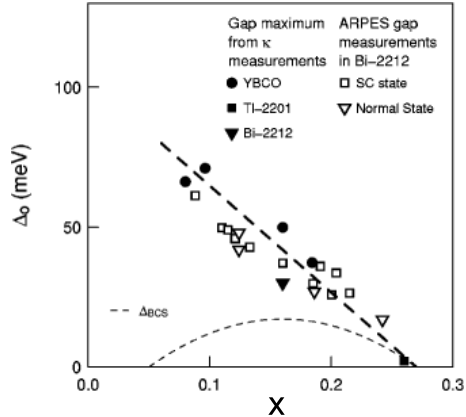}
\caption{Gap amplitude for various cuprates in the pseudogap phase}
\end{figure}

\noindent{\it \bf 5. Concluding Remarks}

The 20th anniversary of the discovery of high-T$_{c}$ cuprate superconductivity is coming
soon.  In the meanwhile, we have learned a lot about the properties of both the normal state and the 
superconducting state in quasi-2D systems and many other systems.  When the effects of 
disorder can be neglected, these
quasiparticles behave as a Fermi liquid although some details may be unorthodox.
Landau's Fermi liquid is thus the most universal normal state in 
high-T$_{c}$ cuprate superconductors, heavy-fermion superconductors and organic conductors.

Furthermore, all the ground states that have been found belong to one of three groups: (a) nodal
superconductors which can be spin singlet or spin triplet; (b) nodal density waves which
can be charge density wave (CDW) or spin density wave (SDW), and (c) the coexistence of
nodal superconductors and nodal density waves.  

The high-T$_{c}$ cuprate superconductors contain all these ground states
in their rich phase diagram.  We are discovering many parallels between high-T$_{c}$
cuprate superconductivity and CeCoIn$_{5}$ \cite{65} 
and $\kappa$-(ET)$_{2}$Cu(N(CN)$_{2}$)Br \cite{66}. 
Unconventional density waves have been identified in 
$\alpha$-(ET)$_{2}$KHg(SCN)$_{4}$
\cite{67} and (TMTSF)$_{2}$PF$_{6}$ \cite{68} through angle dependent magnetoresistance
measurements, and in CeCoIn$_{5}$ \cite{65} and in the pseudogap 
phase of high-T$_{c}$ cuprates \cite{60} through the giant Nernst effect.

In all nodal superconductors, the determination of their gap symmetry is 
the first crucial step.  This allows us to construct the effective Hamiltonian to
describe the plethora of new ground states.

{\bf Acknowledgments}

We thank Thomas Dahm and Peter Thalmeier for early discussions on related subjects and
Balazs Dora, Bojana Korin Hamzi\'{c}, Koichi Izawa, 
Uji Matsuda, Silvia Tomi\'{c}, Andre\'{a} Vanyolos and Attila Virosztek for fruitful collaborations.

HW and KM acknowledge gratefully the hospitality of the Max-Planck Institute for
the Physics of Complex Systems in Dresden, where a part of this work was done.  HW acknowledges
support from the Korean Science and Engineering Foundation (KOSEF) through Grant No.
R05-2004-000-1084.


\begin{references}

\bb{1} J.G. Bednorz and K.A. M\"uller, Z. Physik, B {\bf 64}, 180 (1986).
\bb{2} C.P. Enz, ``A Course in Many-Body Theory applied to Solid State Physics'', 
World Scientific (Singapore 1992).
\bb{3} P.W. Anderson, Science {\bf 235}, 1196 (1987); The Theory of High-T$_{c}$
Superconductivity, Princeton 1998.
\bb{4} J. Bardeen, L.N. Cooper, J.R. Schrieffer, Phys. Rev. {\bf 108}, 1175 (1957).
\bb{5} R.J. Laughlin, cond-mat/0209269.
\bb{6} T. Moriya, Y. Takahasi and K. Ueda, J. Phys. Soc. Jpn. {\bf 59}, 2905 (1990).
\bb{7} D.J. Scalapino, Phys. Rep. {\bf 250}, 329 (1995).
\bb{8} C.H. Pao and N.E. Bickers, PRL {\bf 72}, 1870 (1994).
\bb{9} A. Damascelli, Z. Hussain and Z.X. Shen, Rev. Mod. Phys. {\bf 75}, 473 (2003).
\bb{10} D.J. van Harlingen, Rev. Mod. Phys. {\bf 67}, 515 (1995).
\bb{11} C.C. Tsuei and J.R. Kirtley, Rev. Mod. Phys. {\bf 72}, 909 (2000).
\bb{12} H. Won and K. Maki, Phys. Rev. B {\bf 49}, 1397 (1994).
\bb{13} Y. Sun and K. Maki, Phys. Rev. B {\bf 51}, 6059 (1995).
\bb{14} Y. Sun and K. Maki, Europhys. Lett. {\bf 32}, 355 (1995).
\bb{15} K. Maki and H. Won, Ann. Phys. (Leipzig) {\bf 5}, 320 (1996).
\bb{16} N. Momono et al, Physica C {\bf 264}, 311 (1996).
\bb{17} C. Bernhard et al, Phys. Rev. Lett. {\bf 77}, 2304 (1996).
\bb{18} P.A. Lee, Phys. Rev. Lett. {\bf 71}, 1887 (1993).
\bb{19} May Chiao et al, Phys. Rev. B {\bf 62}, 3554 (2000).
\bb{20} May Chiao, Ph.D thesis, McGill University (1999).
\bb{21} M. Sutherland et al, Phys. Rev. B {\bf 67}, 174520 (2003).
\bb{22} J. Orenstein and A.J. Millis, Science {\bf 288}, 468 (2000).
\bb{23} S. Kivelson et al, Rev. Mod. Phys. {\bf 75}, 1201 (2003).
\bb{24} M. Kato and K. Maki, Europhys. Lett. {\bf 54}, 800 (2001).
\bb{25} M. Kato and K. Maki, Prog. Theor. Phys. (Kyoto) {\bf 107}, 941 (2002).
\bb{26} F. Gygi and M. Schl\"{u}ter, Phys, Rev. B {\bf 41}, 822 (1990); {\bf 43}, 7699 (1991).
\bb{27} M. Franz and Z. Te\^{s}anovi\'{c}, Phys. Rev. Lett. {\bf 80}, 4763 (1998).
\bb{28} K. Yasui and T. Kita, Phys. Rev. Lett. {\bf 80}, 4168 (1999).
\bb{29} M. Takigawa, M. Ichioka and K. Machida, Phys. Rev. Lett. {\bf 83}, 3067 (1999).
\bb{30} M. Ogata, Int. J. Mod. Phys. B {\bf 13}, 3560 (1999).
\bb{31} N.E Hussey, Advances in Physics {\bf 51}, 1685 (2002).
\bb{32} G.E. Volovik, JETP Lett. {\bf 58}, 469 (1993).
\bb{33} I. Vehkter, J. P. Carbotte, E.J. Nicol, Phys. Rev. B {\bf 59}, 7123 (1999).
\bb{34} C. K\"{u}bert and J.P. Hirschfeld, Phys. Rev. Lett. {\bf 80}, 4963 (1998).
\bb{35} H. Won and K. Maki, cond-mat/0004105.
\bb{36} T. Dahm, K. Maki and H. Won, cond-mat/0006301.
\bb{37} H. Won and K. Maki, Curr. Appl. Phys. {\bf 1}, 291 (2001).  Also see H. Won
and K. Maki in ``Vortices in Unconventional Superconductors and Superfluids'', edited
by G.E. Volovik, N. Schopohl and P.R. Huebener (Springer, Berlin 2002).
\bb{38} K. Maki and T. Tsuneto, Prog. Theor. Phys. {\bf 27}, 228 (1962).
\bb{39} K. A. Moler et al, Phys. Rev. Lett. {\bf 73}, 2744 (1994).
\bb{40} B. Revaz et al, Phys. Rev. Lett. {\bf 80}, 3364 (1998).
\bb{41} H. Won and K. Maki, Europhys. Lett. {\bf 56}, 729 (2001).
\bb{42} F. Yu et al, Phys. Rev. Lett. {\bf 74}, 5136 (1993).
\bb{43} H. Aubin et al, Phys. Rev. Lett. {\bf 78}, 2624 (1997).
\bb{44} R. Oca\~{n}a and P. Esquinazi, Phys. Rev. Lett. {\bf 87}, 167006 (2002);
Phys. Rev. B {\bf 66}, 064525 (2002).
\bb{45} H. Won, Q. Yuan, P. Thalmeier and K. Maki, Brazil J. Phys. {\bf 33}, 675 (2003).
\bb{46} K. Maki, S. Haas, D. Parker and H. Won, cond-mat/0407269.   
\bb{47} K. Izawa et al, Phys. Rev. Lett. {\bf 86}, 2653 (2001).
\bb{48} K. Izawa et al, Phys. Rev. Lett. {\bf 87}, 57002 (2001).
\bb{49} K. Izawa et al, Phys. Rev. Lett. {\bf 88}, 27002 (2002).
\bb{50} K. Izawa et al, Phys. Rev. Lett. {\bf 89}, 137006 (2002).
\bb{51} K. Izawa et al, Phys. Rev. Lett. {\bf 90}, 117001 (2003).
\bb{52} L. Benfatto, S. Caprara and C. Di Castro, Eur. Phys. J. B {\bf 17}, 95 (2000).
\bb{53} S. Chakraverty, R.B. Laughlin, D.K. Morr and C. Nayak, Phys. Rev. B {\bf 63},
094503 (2001).
\bb{54} B. Dora, A. Virosztek and K. Maki, Acta Physica Polonica B {\bf 34}, 571 (2003).
\bb{55} B. Dora, K. Maki and A. Virosztek, Modern Phys. Lett. B {\bf 18}, 327 (2004).
\bb{56} T. Timusk and B. Statt, Rep. Prog. Phys. {\bf 62}, 61 (1999).
\bb{57} Y. Wang et al, Phys. Rev. B {\bf 64}, 224519 (2001).
\bb{58} C. Capon et al, Phys. Rev. Lett. {\bf 88}, 056601 (2002).
\bb{59} B. Dora, K. Maki, A. Vanyolos and A. Virosztek, Phys. Rev. B {\bf 68}, 241102 (2003).
\bb{60} K. Maki, B. Dora, A. Virosztek, A. Vangolos, Curr. Appl. Phys., in press.
\bb{61} S. Haas, K. Maki, T. Dahm and P. Thalmeier, cond-mat/0311537.
\bb{62a} B. Dora, A. Virosztek and K. Maki, Phys. Rev. B {\bf 68}, 075104 (2003).
\bb{62} M. Oda et al, Physica C {\bf 282-287}, 1499 (1997).
\bb{63} M. Kugler, O. Fischer, C. Renner, S. Ono and Y. Ando, Phys. Rev. Lett. 
{\bf 86}, 3911 (2001).
\bb{64} B. Dora and A. Virosztek, Euro. Phys. J. B {\bf 22}, 167 (2001).
\bb{65} B. Dora, K. Maki, A. Virosztek, A. Vanyolos, cond-mat/0408351.
\bb{66} M. Pinteri\'{c} et al, Phys. Rev. B {\bf 66}, 174521 (2002).
\bb{67} K. Maki et al, Phys. Rev. Lett. {\bf 90}, 256402 (2003).
\bb{68} B. Dora, K. Maki, A. Vanyolos, A. Virosztek, Europhys. Lett. (in press)



\end{references}
\end{document}